\documentclass[journal=jpcb,manuscript=article]{achemso}

\usepackage[version=3]{mhchem} 
\usepackage{epsfig,amssymb,subfigure}

\newlength{\figwidth}
\author{P. Ziherl}
\affiliation{Faculty of Mathematics and Physics, University of Ljubljana, Jadranska 19, SI-1000 Ljubljana, Slovenia}
\alsoaffiliation{Jo\v zef Stefan Institute, Jamova 39, SI-1000 Ljubljana, Slovenia}
\email{primoz.ziherl@ijs.si}

\author{Randall D. Kamien}
\affiliation{Department of Physics and Astronomy, University of Pennsylvania, 209 S. 33rd St.,
Philadelphia PA 19104-6396, USA}
\email{kamien@physics.upenn.edu}

\title
{From Lumps to Lattices: \\ Crystallized Clots Made Simple} 

\begin{document}

\newpage

\begin{abstract}
Using a minimal model based on the continuum theory of a 2D hard-core/square-shoulder 
ensemble, we reinterpret the main features of cluster mesophases formed by colloids 
with soft shoulder-like repulsive interactions. We rederive the lattice spacing, the binding 
energy and the phase diagram. We also extend the clustering criterion [Likos, C. N., {\sl et al.} 
Phys. Rev. E, {\bf 2001}, {\sl 63}, 031206; Glaser, M. A., {\sl et al.} EPL {\bf 2007}, {\sl 78}, 
46004] to include the effect of the hard cores, which precludes the formation of clusters at 
small densities.
\end{abstract}


\section{Introduction}

Generically, neutral colloidal particles attract each other via van der Waals and Casimir forces, 
and measures must be taken to prevent aggregation of clusters in both experimental systems and 
technological applications. One might be led to conclude that purely attractive hard-core 
particles could only form close-packed lattices or glassy messes. But the phenomenon of 
aggregation in colloids is not restricted to particles that attract each other. Over the past 
two decades clustering in purely repulsive pair potentials has been explored in some detail
to find that it is distinguished by emerging order not seen in attractive particles. In particular, 
it has been established that at large enough density, clumping repulsive colloids can form 
superstructures with large voids on the order of many particle diameters~\cite{Klein94}. 
Indeed, Malescio and Pellicane~\cite{PM03} demonstrated that even simple hard-core, square-shoulder 
potentials led to clustering purely on energetic grounds. The cluster morphologies and the clustering 
criterion itself have been studied theoretically using a range of approaches including 
liquid-state theory~\cite{Likos01}, lattice theory~\cite{Glaser07,Shin09}, density functional 
theory~\cite{Likos07}, and continuum models~\cite{Glaser07,Kosmrlj11}, and the predictions of the 
different approaches are remarkably consistent. Equally unequivocal are the results of numerical 
studies, mostly using Monte Carlo methods~\cite{Mladek06,Glaser07} and direct search of 
minimal-energy configurations using genetic algorithms~\cite{Fornleitner08,Pauschenwein08a,Pauschenwein08b}. 

It is worthwhile at this juncture to step back and develop a minimal set of simple rules to 
expose the mechanism of lattice formation in these systems. It is our hope that these rules will 
make clear the essential ingredients needed for pattern formation in this large class of purely 
repellent systems.  Moreover, we develop these ideas in the context of real-space potentials and 
interactions yet recapitulate the clustering criterion of the more technical (though precise) 
Fourier-based analyses~\cite{Klein94,Likos01,Glaser07}. Designing potentials in real space offers 
a more intuitive route to rational self-assembly and connects directly with, for instance,
laser-trap and depletion based potentials.

In this note, we do this by using the continuum $T=0$ model of the stripe phase formed by particles 
interacting via the hard-core/square-shoulder pair potential
\begin{equation}
U(r)=\left\{\begin{array}{ll}\infty, & r<\sigma\\ \epsilon, & \sigma<r<\lambda\\ 0, & r>\lambda\end{array}\right.
\end{equation}
where $\sigma$ and $\lambda>\sigma$ are the diameters of the core and the shoulder, respectively, and 
$\epsilon$ is the shoulder height~\cite{Kincaid76,Young77}. Characterized by a one-dimensional 
density modulation, the stripe phase is mathematically the most transparent of all cluster 
morphologies and our real-space analysis may be easier to visualize than reciprocal-space 
arguments~\cite{Likos01,Glaser07,Shin09}. Moreover, by focusing on energy rather than on free 
energy we emphasize that the entropy does not promote clustering. These results are then extended
to explore the generic phase diagram of the cluster-forming system at finite temperatures.  

\section{Clusters at $T=0$}

In the following, we first evaluate the energy of a two-dimensional hard-core/square-shoulder system 
with given hard-core diameter $\sigma$ and shoulder diameter $\lambda$ at a fixed average number density, 
and we minimize it with respect to intra-cluster density, cluster size, and lattice spacing. We consider 
the simplest cluster morphology, the one-dimensional square-wave of uniformly populated parallel stripes 
of width $d$ and particle-free gaps of width $\ell-d$; the lattice spacing is $\ell$. In the continuum
model suitable in the large shoulder-to-core limit where $\lambda$ is sufficiently larger than $\sigma$, 
the energy per particle can be related to the average overlap area per particle defined as
\begin{equation}
\omega=\frac{1}{2d}\int_0^d{\rm d}y_1\int{\rm d}{\bf r}_2
\Theta\left(\lambda-|{\bf r}_2-{\bf r}_1|\right),
\label{omega}
\end{equation}
where $\Theta(r)$ is the Heaviside step function describing the shape of the shoulder potential and 
${\bf r}_1$ and ${\bf r}_2$ are the locations of particles 1 and 2, respectively, measured from the 
origin at an edge of the stripe in question. The first integration 
over $y_1$, the distance of particle 1 from a stripe edge, is over the stripe containing particle
1 and the integration over ${\bf r}_2$ goes over all stripes. $\omega$ captures the interaction 
of the shoulders of particles, whereas the hard-core repulsion is treated in a mean-field 
approximation by demanding that the number density within stripes $\rho_{\rm stripes}$ be no larger 
than 
\begin{equation}
\rho_{\rm cp}=\frac{2}{\sqrt{3}\sigma^2}  
\end{equation}
corresponding to the close-packed hexagonal arrangement of the particles' hard-disk cores. This approximation 
is applicable in the broad shoulder regime $\lambda\gg\sigma$ and at densities large enough that the stripes 
are sufficiently wider than the core diameter so that speaking of ``intra-cluster packing'' of particles has 
meaning.

In terms of $\omega$, the average energy per particle reads 
\begin{equation}
E=\epsilon\rho_{\rm stripes}\omega.
\end{equation}
But the number density of particles within stripes depends on their width relative to lattice spacing
$d/\ell$, which represents the fraction of the total area that is occupied by the stripes. In terms
of the average density $\rho$, 
\begin{equation}
\rho_{\rm stripes}=\frac{\rho\ell}{d}.
\end{equation}   
Since the energy is to be minimized at fixed $\rho$ rather than at fixed $\rho_{\rm stripes}$ it is 
convenient to introduce the scaled average overlap area
\begin{equation}
\Omega=\frac{\omega}{d/\ell} 
\end{equation}
which includes all dependence of 
\begin{equation}
E=\epsilon\rho\Omega
\end{equation}
on $d$ and $\ell$. For the hard-core/square-shoulder pair interaction, $\Omega$ can be computed 
analytically but the result is too cumbersome to be of interest here. 

We will now establish some rules of thumb.

\paragraph{I. Clusters are close-packed}

\begin{figure}[h]
\includegraphics{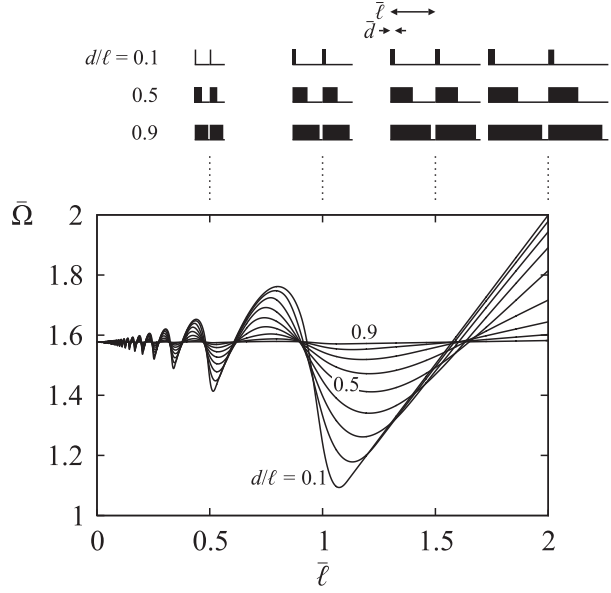}
\caption{Reduced scaled average overlap area as a function of the reduced lattice constant $\bar{\ell}=\ell/\lambda$ 
for $d/\ell=0.1(0.1)0.9$; for the sake of clarity, only curves corresponding to $d/\ell=0.1, 0.5,$ and 0.9 
are labeled. The global minimum of the reduced scaled overlap area $\bar{\Omega}$ at $\bar{\ell}\approx1.2$ is 
deepest for vanishingly small $d/\ell$, which suggests that the equilibrium stripes are as compact as 
possible, {\sl i.e.}, close-packed. The square density waves representing stripes with $d/\ell=0.1,0.5,$ and 
0.9 (top) schematically depict the stripe morphologies at $\bar{\ell}=0.5,1,1.5,$ and $2$; the second column 
where $\bar{\ell}=1$ corresponds to a lattice spacing exactly equal to the shoulder width.}
\label{Omega}
\end{figure}

In \ref{Omega} we plot the reduced scaled average overlap area $\bar{\Omega}=\Omega/\lambda^2$ as a function of 
reduced lattice spacing $\bar{\ell}=\ell/\lambda$ for several stripe widths $d/\ell$. The cluster-free, 
unmodulated phase corresponds to $\bar{\ell}=\bar{d}=0$ and its nature depends on density: As argued below, 
its phase sequence includes the expanded fluid, the expanded hexagonal crystal, the condensed fluid, and the
condensed hexagonal crystal. But since we treat the hard-core part of the pair interaction in a mean-field 
fashion, worrying only about the average number of neighbors within the reach of a particle's shoulder, 
the exact nature of the unmodulated phase is not crucial to understand why clustering takes place.

The uniform, unmodulated phase can be interpreted as a stripe morphology with a very fine density modulation 
such that the lattice spacing and the stripe width are much smaller than the two characteristic length 
scales of the pair potential, $\lambda$ and $\sigma$. In this limit, $\bar{\Omega}=\pi/2$ irrespective of 
$d/\ell$ which tells us that for $d,\ell\ll\lambda$ the average energy per particle is 
$E=\pi\epsilon\rho\lambda^2/2$. As $\bar\ell$ is increased, $\bar{\Omega}$ oscillates around $\pi/2$, 
reaches a global minimum at $\bar{\ell}\approx1.2$, and then grows monotonically to saturate at a 
value of $\pi\ell/2d$. The large $\bar{\ell}$ behavior is a signature of macroscopic phase separation: At
fixed $d/\ell$, states with reduced lattice spacing $\bar{\ell}$ beyond $\sim 1$ correspond to thick 
stripe widths $\bar{d}$ with an ever smaller number of particles residing at the boundary of the stripes. 
The average energy per particle is gradually dominated by that of the particles well within the bulk of 
the stripes, which reads $E=\pi\epsilon\rho_{\rm stripes}\lambda^2/2=\pi\epsilon\rho\lambda^2\ell/2d$ 
because the density within the stripes is larger than the average density by a factor of $\ell/d$. This 
result gives $\bar{\Omega}(\ell\to\infty)=\pi\ell/2d$.

The most important feature of this diagram is that the depth of the minimum decreases with increasing 
relative stripe width $d/\ell$. This means that at any given average density $\rho=\rho_{\rm stripes}d/\ell$, 
the system will select the state with the smallest possible $d/\ell$ at the expense of the density 
within stripes. For example, the curves corresponding to $d/\ell=0.1$ and 0.2 represent two possible 
states of the system of a fixed average density $\rho$. According to \ref{Omega}, the absolute minimum 
of the former is lower than that of the latter, which means that of the two stripe phases in question, the 
$d/\ell=0.1$ state minimizes the total energy $E=\epsilon\rho\lambda^2\bar{\Omega}$. But since the relative 
stripe width of the $d/\ell=0.1$ state is half of that of the $d/\ell=0.2$, the corresponding density
within stripes must be twice as large as in the latter state. In other words, the ground state geometry of 
the stripes minimizes $d/\ell$, restricted only by the close-packing limit forbidding stripes with 
$\rho_{\rm stripes}=\rho\ell/d$ beyond the close-packed density $\rho_{\rm cp}$. We conclude that at $T=0$ 
the optimal stripes are made of close-packed particles so that $\rho_{\rm stripes}=\rho_{\rm cp}$ and the reduced average 
number density 
\begin{equation}
n=\frac{\rho}{\rho_{\rm cp}}
\end{equation}
coincides with $d/\ell$.

\paragraph{II. Lattice spacing weakly depends on density}

An additional feature of the graphs in \ref{Omega} is that the equilibrium reduced lattice spacing, 
$\bar{\ell}_{\rm eq}$, depends only weakly on $d/\ell$. It is apparent that $\bar{\ell}_{\rm eq}$ is 
largest at half-filling where it reaches $1.217$, which nicely agrees with the value of $1.223$ 
predicted by the lattice theory~\cite{Glaser07}. It can be shown that the dependence of 
$\bar{\ell}_{\rm eq}$ on $n$ is well described by a parabola symmetric about $n=0.5$. Even in the 
infinitely-dilute and close-packed limits, $\bar{\ell}_{\rm eq}$ tends to $1$ --- but this result is 
of no physical consequence. As shown below, the binding energy vanishes in these two limits so that
the clustering mechanism is not acting and the corresponding equilibrium reduced lattice spacings are
irrelevant. 

\begin{figure}[h]
\includegraphics{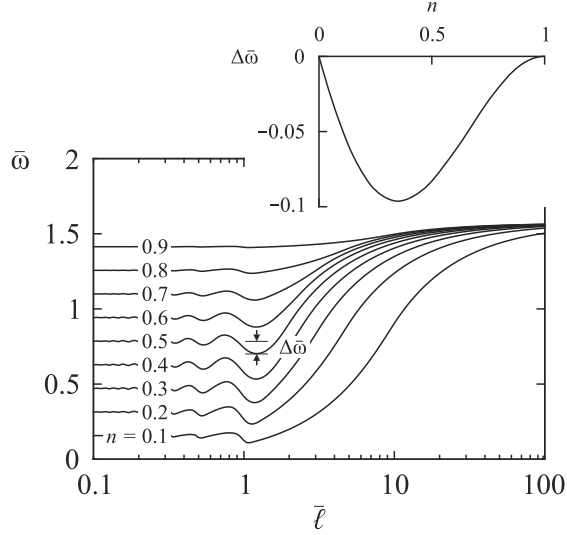}
\caption{Reduced energy of the close-packed stripe morphology, $\bar{\omega}$, as a function of 
reduced lattice spacing $\bar{\ell}$ for $n=0.1(0.1)0.9$. In the unmodulated phase at $\bar{\ell}=0$, 
$\bar{\omega}=n\pi/2$. The inset shows the reduced binding energy 
$\Delta\bar{\omega}=\bar{\omega}(\bar{\ell}_{\rm eq})-\bar{\omega}(\bar{\ell}=0)$ [as illustrated in the 
$\bar{\omega}(n=0.5)$ curve] which is a skewed U-shaped function of reduced average density and vanishes
at $n=0$ and $n=1$.}
\label{omegaf}
\end{figure}

This observation is made even more compelling by employing observation I: Since the ground-state 
stripes are close-packed, we may as well switch back to the average overlap area per particle, 
$\omega$ and write the energy as $E=\epsilon\rho_{\rm cp}\omega$; $\bar{\omega}=\omega/\lambda^2$ can 
thus be regarded as the reduced energy. In \ref{omegaf} we plot $\bar{\omega}$ as a function of 
reduced lattice spacing $\bar\ell$. It is apparent in this presentation that the location of the 
global minimum of $\bar{\omega}$ is little changed over the range $0.1\le n\le 0.9$ and again we 
conclude that $\bar{\ell}_{\rm eq}$ is roughly independent of $n$.

From \ref{omegaf} we also extract the reduced binding energy $\Delta\bar{\omega}$ defined as the depth of the 
global minimum of $\bar{\omega}(n)$ at $\bar{\ell}_{\rm eq}$ relative to the unmodulated phase at $\ell=0$. 
The reduced binding energy shown in the inset to \ref{omegaf} is a skewed U-shaped function of reduced 
average density which vanishes for $n=0$ and $n=1$. In an infinitely dilute system, the energy of
the unmodulated phase itself tends to 0 and no spatial modulation of the density profile can reduce
it further so that the binding energy is 0 too. On the other hand, a system of average density close to
$\rho_{\rm cp}$ cannot undergo but a very restricted spatial modulation (because the density within
the stripes should not exceed $\rho_{\rm cp}$) and thus the energy gained upon clustering approaches 
0 when $\rho\to\rho_{\rm cp}$.

Although $\bar{\omega}$ and $\bar{\Omega}$ are closely related, they convey a somewhat different
message. From \ref{Omega} we learned that the particles within the stripes are close-packed,
which enabled us to directly relate the average density $\rho$ to relative stripe width $d/\ell$. 
On the other hand, \ref{omegaf} exposes the binding energy of the equilibrium stripe morphology 
more clearly. But as far as the magnitude of the equilibrium lattice spacing is concerned, both 
quantities are equally telling. 

\section{Phase diagram}

Our two observations can be used to qualitatively outline the phase diagram of the cluster phases 
in the temperature-density plane. To this end, we need to study the difference of the free energies
of the stripe phase and the unmodulated phase, which consists of an energy term 
$\Delta E=\epsilon\rho_{\rm cp}\lambda^2\Delta\bar{\omega}$ and of an entropic term. The entropy of 
the two phases depends on their structure elaborated below in a semi-quantitative fashion.

At absolute zero, one of the hallmark features of the stripe morphology is its compact intra-stripe 
structure where the impenetrable hard cores of the particles are packed together as tightly as 
possible at any reduced average density $n$. This state is materialized by the hexagonal lattice. Because of
the robust, density-independent nature of this behavior, we posit that at finite temperatures 
the intra-stripe density $\rho_{\rm stripes}$ should not depend strongly on the reduced average 
density either (though it must be smaller than the close-packing density $\rho_{\rm cp}$) and that the intra-stripe order 
remains hexagonal. 

\begin{figure}[ht]
\includegraphics{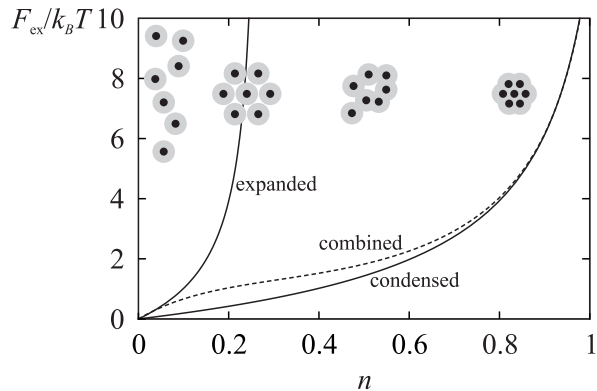}
\caption{Excess entropic free energies per particle of the expanded and the condensed unmodulated phases derived 
using a Carnahan-Starling type theory (solid lines; the expanded phase corresponds to $\lambda/\sigma=2$). 
The divergence of $F_{\rm ex}^{\rm exp}$ at $n=0.276$ is unphysical; instead of approaching the 
shoulder-to-shoulder close-packed structure, the system melts to form the condensed fluid phase. The 
dashed line represents a qualitatively correct interpolation between the expanded and the condensed 
branch. --- The schematics illustrate the structure of the 4 unmodulated phases: The expanded fluid,
the expanded crystal, the condensed fluid, and the condensed crystal. Full circles indicate the hard 
cores of particles whereas the shaded coronas represent the shoulders.}
\label{Fex}
\end{figure}

The structure of the unmodulated phases of hard-core/square-shoulder particles is more complicated 
and despite decades of efforts (see, {\sl e.g.}, Refs.~\cite{Rascon97,Lang99,Singh10}), their thermodynamics 
remains only a partly solved problem. For the purpose of present discussion, it suffices to note that at 
low temperatures where they compete with cluster morphologies, the phase sequence of unmodulated 
phases consists of 4 variants, the 2 low-density and high-density phases being characterized 
with little and sizable overlap of particles' shoulders, respectively. At very small densities, 
the particles form an expanded fluid of disks of shoulder diameter 
$\lambda$ (schematically shown in \ref{Fex}). As this fluid is compressed, it undergoes a transition to 
the expanded hexagonal crystal of disks of diameter $\lambda$. The location of the transition can be 
estimated by rescaling the phase diagram of the hard-disk system: $\rho_{\rm ef-ec}\approx0.792(\sigma/\lambda)^2
\rho_{\rm cp}$~\cite{Binder02}. Upon further compression, the expanded hexagonal phase remelts to avoid 
close-packing; since the overlap of shoulders is increasingly less unfavorable at elevated temperatures, 
the phase transition density should decrease with temperature. From this transition on, the particles 
behave essentially as hard spheres of diameter $\sigma$ and the transition to the condensed hexagonal 
phase takes place approximately at $\rho_{\rm cf-cc}\approx0.792\rho_{\rm cp}$~\cite{Binder02}.

Understanding the main features of the sequence of unmodulated phases helps us to construct 
semi-qualitatively the entropic part of their free energy. We first note that the pressure of the hard-disk 
hexagonal crystal can be roughly regarded as a continuation of the hard-disk fluid branch~\cite{Binder02} 
and we approximate the excess entropic free energy per particle in both the fluid and the crystal 
unmodulated phase by
\begin{equation}
F_{\rm ex}^{\rm cond}(n)=k_BT\left[\frac{\alpha n}{1-\alpha n}-\ln(1-\alpha n)\right]
\label{Fexe}
\end{equation}
where $\alpha=\pi/2\sqrt{3}\approx0.907$. This prediction is based on the 2D Carnahan-Starling type theory 
for the fluid phase~\cite{Maeso93,Hansen86}. Admittedly a rough approximation --- it does not 
distinguish between the fluid and the crystalline phase, and the hexatic phase is disregarded altogether 
--- this excess entropic free energy provides a simple and adequate model across a broad range of reduced 
densities consistent with the scope of this analysis. Its main deficiency is the poor description of the
crystalline phase (whose excess entropic free energy should diverge at reduced average density of $n=1$ 
and not at $n=1/\alpha=1.103$). Yet we note that in the phase diagram of hard disks, the condensed 
crystalline phase is stable at $n>0.792$~\cite{Binder02} so that the discrepancy is limited only to large 
densities. 

Using the model $F_{\rm ex}$ [Eq.~(\ref{Fexe})], we can outline the excess entropic free energy for the 
unmodulated phases. In \ref{Fex}, we plot the excess free energy per particle of the expanded and the condensed phases, 
$F_{\rm ex}^{\rm exp}(n)=F_{\rm ex}^{\rm cond}\left((\lambda/\sigma)^2n\right)$ and $F_{\rm ex}^{\rm cond}(n)$. 
The former diverges at $n=(\sigma/\lambda)^2/\alpha$ which corresponds to close-packed disks of diameter
$\lambda$. This divergence is, of course, unphysical because the expanded hexagonal lattice remelts upon
compression so that in this regime, the true excess entropic free energy interpolates between 
$F_{\rm ex}^{\rm exp}$ and $F_{\rm ex}^{\rm cond}$.

The exact shape of the interpolating $F_{\rm ex}$ is not known. But our mean-field model is designed
to work best for broad shoulders and in this case, the existence of the expanded phases is restricted
to reduced average densities below $(\sigma/\lambda)^2/\alpha$, {\sl i.e.}, to very small $n$. Being interested
in the overall behavior of the system, we may approximate the excess entropic free energy by the
condensed branch alone. Then the total free energy difference can be constructed from overlap energy 
difference $\Delta E(n)$, the excess entropic free energy of the stripe morphology 
$F_{\rm ex}^{\rm cond}(n_{\rm stripes})$, and the excess entropic free energy of the unmodulated phase 
$F_{\rm ex}^{\rm cond}(n)$:
\begin{eqnarray}
\Delta F(n,n_{\rm stripes},\tau)&=&\Delta E(n)+F_{\rm ex}^{\rm cond}(n_{\rm stripes})-F_{\rm ex}^{\rm cond}(n)\nonumber \\
&=&\epsilon\rho_{\rm cp}\lambda^2\Bigg\{n_{\rm stripes}\Delta\bar{\omega}(n)\nonumber\\
&&+\tau\left[
\frac{\alpha n_{\rm stripes}}{1-\alpha n_{\rm stripes}}-\ln(1-\alpha n_{\rm stripes})
-\frac{\alpha n}{1-\alpha n}+\ln(1-\alpha n)\right]\Bigg\}
\end{eqnarray}
Here $n_{\rm stripes}=\rho_{\rm stripes}/\rho_{\rm cp}$ is the reduced density within stripes, 
which should not depart much from $1$ and must decrease with temperature, and 
\begin{equation}
\tau=\frac{k_BT}{\epsilon\rho_{\rm cp}\lambda^2}
\end{equation}
is the reduced temperature.

\paragraph{The clustering criterion}

The most important features of the total free energy difference are the negative skewed U-shaped energy term, 
whose exact dependence on the reduced average density shown in the inset to \ref{omegaf} can be well fitted by
\begin{equation} 
\Delta E\approx-0.65\epsilon\rho_{\rm cp}\lambda^2n(1-n)^2, 
\end{equation}
and the positive excess entropic free energy difference proportional to temperature which monotonically decreases 
from a finite value at $n=0$ to 0 at $n=n_{\rm stripes}$. We approximate the temperature dependence of the 
reduced density within stripes by a linearly decreasing function
\begin{equation}
n_{\rm stripes}(\tau)=1-c\tau.
\label{nstr}
\end{equation}
In the following, we choose $c=6$: At the largest reduced temperature where the stripe morphology is stable, this 
gives $n_{\rm stripes}\approx0.9$ which is plausible.  

\begin{figure}[hp]
\includegraphics{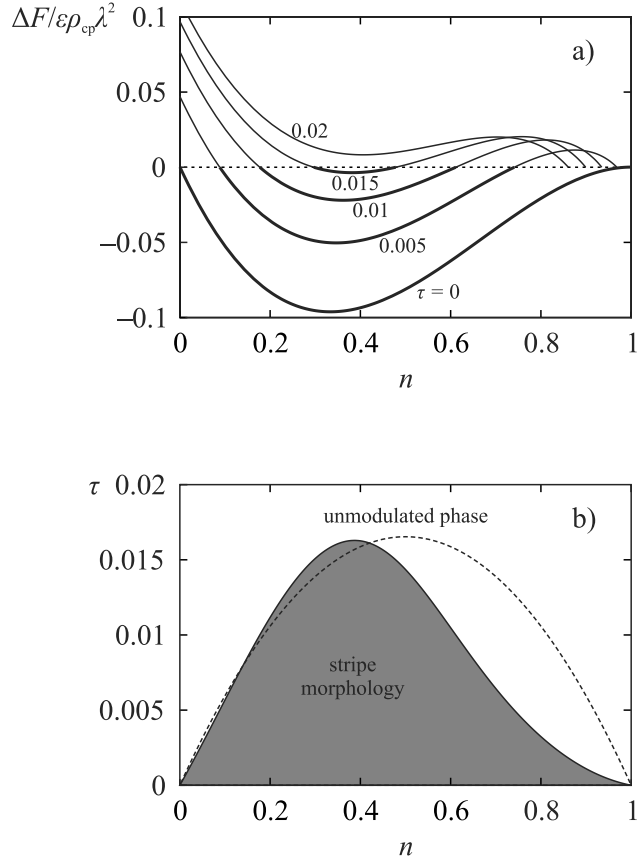}
\caption{Free energy difference of the stripe and the unmodulated phase for $n_{\rm stripes}$ described by
Eq.~(\ref{nstr}) and reduced temperature $\tau=0,0.005,0.01,0.015,$ and $0.02$ (panel a). As temperature 
is increased, the range of reduced average densities $n$ where $\Delta F<0$ and the stripe morphology is 
stable (thick sections of curves) becomes increasingly more narrow. Each curve terminates at $n^*\lesssim 
n_{\rm stripes}(\tau)$ such that $\Delta F(n>n^*)<0$; states beyond this point correspond to average density 
very similar to the density within stripes where the predictions of our model are meaningless. --- Panel b) 
shows the temperature-density phase diagram of the hard-core/soft-shoulder stripe morphology and the unmodulated 
phase computed using the mean-field continuum model with the model $n_{\rm stripes}(\tau)$ [Eq.~(\ref{nstr})]. The 
stripe morphology is stable in the shaded region whose shape agrees rather well with the clustering criterion 
[dashed line; $const.$ in Eq.~(\ref{clustering}) adjusted to reproduce the slope of the phase boundary at small 
reduced average densities].}
\label{phased}
\end{figure}

In \ref{phased}a, we plot $\Delta F$ for several values of reduced temperature $\tau$. As $\tau$ is increased, 
the reduced average density range where the stripe morphology is stable gradually shrinks and at a large enough 
$\tau$, the stripe morphology is disfavored at any $n$. Thus the phase diagram is characterized by a dome-like region 
of stability of the stripe morphology shown in \ref{phased}b. Except at very large $n$ where the differences between 
the unmodulated and the stripe phase are increasingly smaller and our model is inaccurate, the shape of the 
phase boundary reproduces well the clustering criterion obtained in terms of the lattice theory~\cite{Glaser07}
which states that instability occurs when 
\begin{equation}
\frac{n(1-n)}{\tau}>const.>0.
\label{clustering}
\end{equation}
The agreement is indeed remarkable although our phase diagram lacks the symmetry about the half-filling point 
$n=1/2$ encoded in Eq.~(\ref{clustering}). Needless to say, the model can be refined by better estimating the 
entropic part of the excess free energy. 
  
\paragraph{Extending the clustering criterion}

The stripe morphology is subject to two consistency constraints. Firstly, adjacent stripes separated 
by more than a shoulder width do not interact with each other because there is no overlap of particles
residing within them; without a restoring interstripe repulsion, such configurations would spontaneously 
disintegrate into thinner stripes with narrower gaps between them. This means that in the mechanically 
stable stripe state, the width of the particle-free gaps between stripes $\ell-d$ should be smaller than 
the shoulder diameter $\lambda$. A close inspection of \ref{Omega} shows that the global minimum at 
reduced lattice spacing $\bar{\ell}\approx1$ is not compromised by this condition at any reduced average 
density $n$. Secondly, the stripe width must be larger than the hard-core diameter of the particles, 
$d>\sigma$; if not, speaking of close-packed stripes does not make sense. These two conditions bracket 
the physically relevant range of lattice spacing $\ell$ from top and from bottom, respectively. To compare 
the two bounds, we replace $d$ by $n\ell$ so that i) the upper bound $\ell-d<\lambda$ becomes 
$\ell(1-n)<\lambda$ wherefrom $\ell<\lambda/(1-n)$ and ii) the lower bound $d>\sigma$ becomes $n\ell>\sigma$ 
and thus $\ell>\sigma/n$.  We thus find that a stable phase can only exist if 
\begin{equation}
n>\frac{\sigma}{\lambda+\sigma},
\label{restriction}
\end{equation} 
the lower limit of stability of the stripe morphology. In view of the nature of the continuum model used here, 
this treatment of the hard-core part of the pair potential is expected to be valid for core-to-shoulder 
ratios $\lambda/\sigma$ sufficiently smaller than 1.

Thus the effect of the hard-core part of the pair potential is to disfavor clustering at small densities, 
thereby restricting the validity of the criterion [Eq.~(\ref{clustering})] to densities beyond a threshold 
determined by the core-to-shoulder ratio. Since the hard-core interaction is athermal, this condition 
should apply at all $T$ as depicted in \ref{phased2}.

\section{Conclusions}

\begin{figure}[h]
\includegraphics{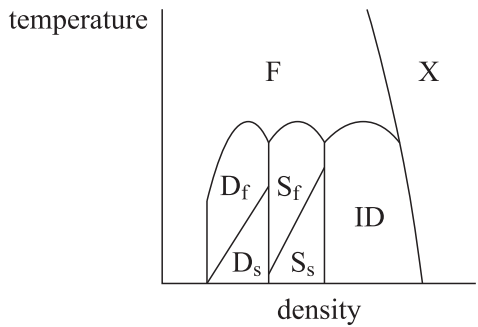}
\caption{Generic phase diagram of cluster-forming repelling particles. At low temperatures, the phase 
sequence includes the fluid phase (F), several cluster morphologies (fluid disks --- $\rm D_f$,
solid disks --- $\rm D_s$, fluid stripes --- $\rm S_f$, solid stripes ---  $\rm S_s$, and
inverted disks --- ID), and one or more crystal phases (X). The vertical boundary of the disk morphologies
at small densities indicates the restriction imposed by the hard-core part of the pair potential 
[Eq. (\ref{restriction})]. Regions of phase coexistence are not shown for clarity.}
\label{phased2}
\end{figure}

The complete phase diagram will include the fluid and one or more crystal lattices as the low- and
the high-density variants of the unmodulated, non-cluster phase. On top of the stripe morphology, in two 
dimensions there also exist the disk and the inverted disk cluster phase~\cite{Glaser07}. Just like stripes 
are most stable at about $n=0.4$ (\ref{phased}) which corresponds to $\bar{\ell}\approx1.2$ and $\bar{d}=0.48$, 
the disks are expected to be bound most tightly at a similar lattice spacing and disk diameter. This 
configuration of disks will cover a smaller fraction of the plane and the corresponding average density 
will be smaller than $0.4$. If we assume that the disks are stable in a dome-like region of the phase 
diagram qualitatively similar to that describing the stripes, the disk dome must peak at a density 
smaller than that of the stripe dome. Conversely, the inverted disk morphology should prevail at densities 
larger than $0.4$. Moreover, the disk and the stripe phases occur in both the liquid and the solid 
intra-cluster order~\cite{Glaser07} and so the generic full phase diagram of a cluster-forming ensemble 
should have a multiple-dome structure curbed by the fluid phase at small densities and by the crystal 
phases at large densities (\ref{phased2}). At large temperatures, the fluid-crystal phase transition line 
must be vertical because for $T\to\infty$, the system reduces to hard disks of diameter $\sigma$ which 
freeze and melt at $\rho/\rho_{\rm cp}=0.779$ and $0.792$, respectively~\cite{Binder02}.

\ref{phased2} reproduces many features of the phase diagram obtained using a more complete treatment of the 
thermodynamics of a hard-core/square-shoulder system~\cite{Glaser07} and it bears some similarity to the 
phase diagram of the hard-core/linear-ramp system~\cite{Jagla98}. Although the width of the ramp studied in 
Ref.~\cite{Jagla98} is too narrow for fully developed cluster phases, the non-close-packed lattices occurring 
in the phase diagram are very reminiscent of the cluster morphologies discussed here, and the phase diagram itself 
has roughly the same multidome shape as that in \ref{phased2}. We expect that for the hard-core/square-shoulder 
potential with small core-to-shoulder ratio, the agreement of the numerically obtained phase diagram with our 
prediction should be even better.

The ideas presented here capture the main mechanisms of cluster formation in systems of classical repelling 
particles in a way marked by the appeal of real-space description and by the analysis of the density-modulated 
morphologies across the whole range of lattice spacing. Given the seemingly counterintuitive behavior of particles 
with shoulder-type pair interaction, we hope that our rederivation will clarify the details of the more elaborate 
studies.  

\subsection{Acknowledgment}

We thank H. Diamant, M. A. Glaser, G. M. Grason, C. J. Olson Reichhardt, C. Reichhardt, and C. D. Santangelo for 
helpful discussions and we acknowledge the hospitality of the Aspen Center for Physics where a part of this work was 
done. This work was supported by the Marie-Curie Initial Training Network COMPLOIDS under FP7-PEOPLE-ITN-2008 Grant 
No.~234810. PZ was supported by Slovenian Research Agency through Grant No.~P1-0055. RDK was supported through NSF Grant 
DMR05-47230.

\newpage

\begin{figure}[p]
\includegraphics[width=50mm]{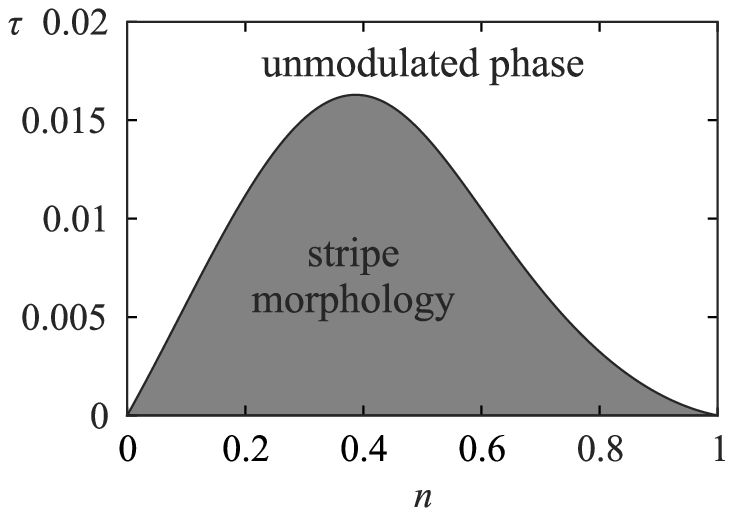}
\caption{Ziherl and Kamien: Graphic for the Table of contents}
\end{figure}


\begin{thebibliography}{100}
\bibitem{Klein94} Klein, W.; Gould, H.; Ramos, R. A.; Clejan, I.; Mel'cuk, A. I. Physica A {\bf 1994},
{\sl 205}, 738-746. 
\bibitem{PM03} Malescio, G.; Pellicane, G. Nat. Mater. {\bf 2003}, {\sl 2}, 97-100.
\bibitem{Likos01} Likos, C. N.; Lang, A.; Watzlawek, M.; L\"owen, H. Phys. Rev. E
{\bf 2001}, {\sl 63}, 031206.
\bibitem{Shin09} Shin, H.; Grason, G. M.; Santangelo, C. D. Soft Matter {\bf 2009}, {\sl 5}, 3629-3638.
\bibitem{Glaser07} Glaser, M. A.; Grason, G. M.; Kamien, R. D.; Ko\v smrlj, A.; Santangelo, C. D.; 
Ziherl, P. EPL {\bf 2007}, {\sl 78}, 46004.
\bibitem{Likos07} Likos, C. N.; Mladek, B. M.; Gottwald, D.; Kahl, G. J. Chem. Phys. {\bf 2007},
{\sl 126}, 224502.
\bibitem{Kosmrlj11} Ko\v smrlj, A.; Pauschenwein, G. J.; Kahl, G.; Ziherl, P. to be published.
\bibitem{Mladek06} Mladek, B. M.; Gottwald, D.; Kahl, G.; Neumann, M.; Likos, C. N. Phys. Rev. Lett. 
{\bf 2006}, {\sl 96}, 045701.
\bibitem{Fornleitner08} Fornleitner, J.; Kahl, G. EPL {\bf 2008}, {\sl 82}, 18001.
\bibitem{Pauschenwein08a} Pauschenwein, G. J.; Kahl, G. Soft Matter {\bf 2008}, {\sl 4}, 1396-1399. 
\bibitem{Pauschenwein08b} Pauschenwein, G. J.; Kahl, G. J. Chem. Phys. {\bf 2008}, {\sl 129}, 174107.
\bibitem{Kincaid76} Kincaid, J. M.; Stell, G.; Goldmark, E. J. Chem. Phys. {\bf 1976}, {\sl 65}, 2172-2179.
\bibitem{Young77} Young, D. A.; Alder, B. J. Phys. Rev. Lett. {\bf 1977}, {\sl 38}, 1213-1216.
\bibitem{Kirkwood} Kirkwood, J. G. J. Chem Phys. {\bf 1950}, {\sl 18}, 380-382.
\bibitem{Rascon97} Rasc\' on, C.; Velasco, E.; Mederos. L.; Navascu\' es, G. J. Phys. Chem. {\bf 1997},
{\sl 106}, 6689-6697.
\bibitem{Lang99} Lang, A.; Kahl, G.; Likos, C. N.; L\"owen, H.; Watzlawek, M. J. Phys. Condens. Mat.
{\bf 1999}, {\sl 11}, 10143-10161.
\bibitem{Singh10} Singh, M.; Liu, H.; Kumar, S. K.; Ganguly, A.; Chakravarty, C. J. Chem. Phys. {\bf 2010},
{\sl 132}, 074503.
\bibitem{CS} Carnahan, N. F.; Starling, K. E. J. Chem. Phys. {\bf 1967}, {\sl 51}, 635-636.
\bibitem{Jagla98} Jagla, E. A. Phys. Rev. E {\bf 1998}, {\sl 58}, 1478-1486. 
\bibitem{Binder02} Binder, K.; Sengupta, S.; Nielaba, P. J. Phys. Condens. Mat. {\bf 2002}, {\sl 14},
2323-2333.
\bibitem{Maeso93} Maeso, M. J.; Solana, J. R. J. Chem. Phys. {\bf 1993}, {\sl 99}, 548-552.
\bibitem{Hansen86} Hansen, J.-P.; McDonald, I. R. {\sl Theory of Simple Liquids} (Academic Press, 
London, 1986). 
\end{thebibliography}
\end{document}